\documentclass[5p, twocolumn,10pt,sort&compress]{elsarticle} 

\usepackage{url}
\usepackage{amssymb}
\usepackage{graphicx}
\usepackage{float}
\usepackage{amsmath}
\usepackage{xcolor}
\usepackage{soul}

\journal{Physics Letters B}
\begin{document}
\begin{frontmatter}

\title{Trans-Planckian censorship constraints on properties and \\cosmological applications of axion-like fields}
\author{David Shlivko}
\ead{dshlivko@princeton.edu}
\affiliation{organization={Department of Physics, Princeton University},
            city={Princeton},
            state={NJ 08544},
            country={USA}}

\begin{abstract}
We use the Trans-Planckian Censorship Conjecture (TCC) to constrain the decay constants $f$ characterizing a set of $\mathcal{N}$ identical axion-like fields with cosine potentials, improving upon the precision of other Swampland conjectures and existing string-theoretic arguments. We find that consistency with the TCC requires any such set of axion-like fields to satisfy  $f\sqrt{\mathcal{N}} \lesssim 0.6M_{pl}$, where $M_{pl}$ is the reduced Planck mass. We show that this bound makes models of axion-driven inflation incapable of simultaneously producing the required number of e-foldings and the observed scalar spectral tilt. In contrast, we find that models of axion quintessence can be simultaneously compatible with the TCC and observational data, provided that the axions' initial field values are set near the maxima of their potentials to within roughly $\pm \frac{\pi}{5}f$.
\end{abstract}


\end{frontmatter}

\section{Introduction}\label{sec:intro}
In the context of effective field theories, an axion-like field (henceforth, ``axion'') is a pseudoscalar angular degree of freedom that emerges as the Nambu-Goldstone mode of a complex scalar field with a spontaneously broken chiral U(1) symmetry. The axion's continuous shift symmetry can be explicitly broken at sufficiently low energies by non-perturbative effects, such as instanton transitions within a non-Abelian gauge sector coupled to the axion. These effects give the axion $\varphi$ a periodic effective potential that takes the form $V_{\text{eff}} \propto \cos(\varphi/f)$ in the dilute instanton gas approximation, up to an additive constant \cite{callan_toward_1978}. Here the decay constant $f$ is set by the scale of spontaneous symmetry breaking. The scope of this work will be restricted to potentials of this form. Except for a brief discussion related to axion quintessence, we will also assume that the net vacuum energy is negligible compared to the amplitude of the axion potential, setting 
\begin{equation}\label{cosV}
	V(\varphi) = m^2f^2[\cos(\varphi/f)+1],
\end{equation} 
with  $m = |V''(n\pi f)|, \; n \in \mathbb{Z},$ denoting the axion's nominal mass. The mass is exponentially sensitive to the instanton action and can therefore take on a wide range of values depending on the gauge sector to which the axion is coupled \cite{callan_toward_1978}. Moreover, the axion's mass is shielded from large perturbative loop corrections due to its (weakly broken) shift symmetry \cite{frieman_cosmology_1995}. These factors have motivated the use of axions as candidates for a wide range of cosmological models postulating the existence of spin-zero fields with low masses and flat potentials, including models of inflation and quintessence (see, e.g., \cite{marsh_axion_2016} for a review). 

In addition to being phenomenologically interesting, axions are on strong theoretical footing due to their generic emergence in the low-energy limit of string theory \cite{svrcek_axions_2006}. These ``string axions'' can acquire a similarly wide range of masses from worldsheet or membrane instantons \cite{witten_properties_1984, svrcek_axions_2006}, though contributions from other non-perturbative effects remain uncertain \cite{conlon_qcd_2006, demirtas_pq_2022}. Despite this uncertainty, attempts to model the accelerated expansion of the universe using string axions have continued for decades \cite{choi_string_1999, svrcek_cosmological_2006, grimm_axion_2007, panda_axions_2011, gupta_observational_2011, kamionkowski_dark_2014, cicoli_n-flation_2014}.

Independently, string-theoretic arguments (and considerations of quantum gravity more broadly) have led to the development of Swampland conjectures that place limits on viable effective field theories and on the possible dynamics of cosmic expansion \cite{vafa_string_2018, garg_bounds_2019, ooguri_distance_2019, bedroya_trans-planckian_2020}. One such conjecture is the Trans-Planckian Censorship Conjecture (TCC), which states that a phase of accelerated expansion will never last long enough for sub-Planckian perturbation modes to be stretched to super-Hubble length scales \cite{bedroya_trans-planckian_2020}. In addition to being consistent with explicit constructions in string theory, the TCC has been supported by general arguments from holography and gravitational renormalizability, and it is connected to many of the other Swampland conjectures \cite{bedroya_holographic_2022, bedroya_sitter_2021, heisteeg_bounds_2023}. As a result, the scope of the TCC extends to systems containing axions of any origin compatible with quantum gravity, whether string-theoretic or otherwise.

In this work, we analyze the implications of the TCC for axions in general and for models of axion-driven cosmic acceleration in particular.
Our central goal will be to place constraints on the decay constants characterizing systems of $\mathcal{N}$ identical axions with masses $m \ll M_{pl}$, where $M_{pl} = \sqrt{\hbar c / (8\pi G)}$ is the reduced Planck mass. $\mathcal{N}$-axion systems are frequently used in models of cosmic acceleration \cite{kaloper_pngb_2006,  svrcek_cosmological_2006, dimopoulos_n-flation_2008, kim_n-flation_2006, easther_random_2006, olsson_inflation_2007, cicoli_n-flation_2014}, motivated by the failure of string theory to produce an axion with a sufficiently large decay constant ($f \gtrsim M_{pl}$) to drive single-field inflation \cite{freese_natural_1990, adams_natural_1993, banks_possibility_2003} or quintessence \cite{frieman_cosmology_1995, kaloper_pngb_2006, svrcek_cosmological_2006}. 
The goal of these models is typically to use many axions with small decay constants to mimic a single axion with a large ``effective'' decay constant, $f_\text{eff} \equiv f\sqrt{\mathcal{N}}$, that could drive accelerated expansion \cite{dimopoulos_n-flation_2008, kaloper_pngb_2006}. 
Later arguments from string theory and the Swampland program, however, have suggested that even $f_\text{eff}$ is bounded from above by the Planck scale \cite{heidenreich_weak_2015, reig_stochastic_2021, agrawal_cosmological_2018, rudelius_possibility_2015}. 

In Section \ref{sec:tcc}, we improve upon the precision of these bounds by showing that $\mathcal{N}$-axion systems are only compatible with the TCC if $f_\text{eff} \lesssim 0.6 M_{pl}$. In Section \ref{sec:inflation}, we show that this bound rules out models of $\mathcal{N}$-axion inflation to a much greater degree than the existing $2\sigma$ tensions between measurements of the cosmic microwave background (CMB) and predictions from axion-driven ``natural inflation'' scenarios \cite{akrami_planck_2020}. On the other hand, we find in Section \ref{sec:quintessence} that the TCC is compatible with $\mathcal{N}$-axion quintessence models, as long as the initial field values lie sufficiently close to the maxima of the axions' cosine potentials. Finally, in Section \ref{sec:conclusions}, we summarize our results, elaborate on the different implications for quintessence models when the axions are string-theoretic vs. non-string-theoretic in origin, and comment on possible extensions of this study.
Note that throughout the remainder of this work, we will work in units where $M_{pl} = 1$.

\section{TCC Constraints on Axions}\label{sec:tcc}
The TCC restricts the duration of any phase of accelerated expansion of the universe by forbidding sub-Planckian perturbation modes from being stretched to super-Hubble length scales. Mathematically, this statement can be written as
\begin{equation}\label{tcc}
	\frac{a_\text{end}}{a_0} \leq \frac{M_{pl}}{H_\text{end}},
\end{equation}
where $a_0 \equiv 1$ is the scale factor at the onset of accelerated expansion, and $a_\text{end}$ and $H_\text{end}$ are the scale factor and Hubble parameter at its completion. This restriction can be used to constrain the potential of any scalar field by requiring that Eq.  (\ref{tcc}) is obeyed throughout the field's evolution for \emph{all} physically allowed initial conditions \cite{bedroya_trans-planckian_2020}. Note that where quantum fluctuations or tunneling events are concerned, the TCC has been interpreted as a probabilistic statement, with Eq. (\ref{tcc}) holding for the expected amplitude of fluctuations or expected tunneling time \cite{bedroya_trans-planckian_2020}.

In this section, we will first develop an approximate analytic constraint on the effective decay constant $f_\text{eff} \equiv f\sqrt{\mathcal{N}}$ characterizing the potential of an $\mathcal{N}$-axion system, and we will then tighten this constraint using more accurate numerical methods. For the analytic calculation, we choose to give each field $\varphi_n$ (where $n = 1, \, ... \, ,\, \mathcal{N}$) an initial value 
\begin{equation}\label{phi0}
	\varphi_n(0) = \frac{mf_\text{eff}^2}{\pi\sqrt{6}}
\end{equation} 
and initial velocity 
\begin{equation}\label{phid0}
	\dot{\varphi}_n(0) = \frac{m^2f_\text{eff}}{6\pi}
\end{equation}
at some initial time $t = 0$. We also restrict our analysis to axions with masses satisfying  $m \ll 1/(\mathcal{N}f)$ or $m \ll 1$, whichever is stricter. This ensures that $\varphi_n(0) \lesssim f$ and $\dot{\varphi}_n(0)^2 \ll V(\varphi_n(0))$, allowing us to approximate
\begin{equation}\label{approxVp}
	V'(\varphi_n) \approx -m^2\varphi_n
\end{equation}
and, per the Friedmann Equation,
\begin{equation}\label{approxH}
	3H^2 \approx \sum_{n=1}^N V(\varphi_n) \approx 2m^2f_\text{eff}^2.
\end{equation}

 Our choice of initial conditions---which, no matter how contrived, must still obey the TCC---places the fields sufficiently close to the hilltop ($\varphi_n = 0$) to drive accelerated expansion, but sufficiently far from the hilltop (and with sufficiently large velocities) to make quantum fluctuations subdominant to classical evolution. We derive this latter statement in \ref{app}.

Using the initial conditions (\ref{phi0}-\ref{phid0}) and the approximations (\ref{approxVp}-\ref{approxH}), we solve the equations of motion
\begin{equation}\label{eom}
	\ddot{\varphi}_n + 3H\dot{\varphi}_n + V'(\varphi_n) = 0
\end{equation}
to find an exponential growing mode
\begin{equation}\label{traj_top}
	\varphi_+(t) = Ae^{\omega_+t}, 
\end{equation}
where
\begin{align}
	A &\equiv \varphi_n(0) \cdot \frac{3+f_\text{eff}^{-2}+\sqrt{9+6f_\text{eff}^{-2}}}{2\sqrt{9+6f_\text{eff}^{-2}}}, \\
	\omega_+ &\equiv \frac{H}{2}\left(\sqrt{9+6f_\text{eff}^{-2}}-3\right).\label{omega}
\end{align}
Note that $A \approx \varphi_n(0)$ for any $f_\text{eff} \gtrsim 0.1$, and the inverse time constant is $\omega_+ \approx H/(2f_\text{eff}^2)$ in the limit $f_\text{eff} \gg 1$ or alternatively $\omega_+ \approx m$ in the limit $f_\text{eff} \ll 1$. 

Accelerated expansion continues at least until $\varphi_n \sim f$, where one also has from Eq. (\ref{traj_top}) that $\dot{\varphi}_n \sim \omega_+f$. This can be verified by computing the equation of state, $\epsilon \equiv \frac{3}{2}(1+w) \equiv \frac{3}{2}(1 + P/\rho)$, where $P$ and $\rho$ are respectively the total pressure and energy density. In particular, for a collection of identical scalar fields, we have that
\begin{equation}\label{epsilon}
	\epsilon = \frac{3\dot{\varphi}_n^2}{\dot{\varphi}_n^2+2V(\varphi_n)} 
	\approx \frac{3}{1 + 3(m/\omega_+)^2(f/\varphi_n)^2}.
\end{equation}
When $f_\text{eff} \ll 1$, we see that $\epsilon \approx \frac{3}{1 + 3(f/\varphi_n)^2}$ approaches $\mathcal{O}(1)$ as $\varphi_n \rightarrow f$, signaling the end of acceleration of the scale factor (which obeys $\ddot{a} \propto (1 - \epsilon)$). In the opposite limit ($f_\text{eff} \gtrsim 1$), the equation of state is still well below unity when $\varphi_n \approx f$, meaning acceleration will go on for a bit longer---but it will certainly end before the fields reach the minima at $\varphi_n = \pi f$, where $V = 0$ and $\epsilon = 3$. Therefore, in either case, we can conservatively assume that acceleration lasts from $\varphi_n \sim \varphi_n(0)$ until $\varphi_n \sim f$, and the scale factor at the end of acceleration satisfies
\begin{equation}\label{nend}
	\ln(a_\text{end}) = \int_{t=0}^{t_\text{end}} H(t)dt \gtrsim H_\text{end}t_\text{end} \approx \frac{H_\text{end}}{\omega_+} \ln\left(\frac{f}{A}\right).
\end{equation}
The Friedmann equation gives
\begin{equation}
	H_\text{end} = \sqrt{\frac{\rho_\text{end}}{3}} \geq \sqrt{\frac{V_\text{end}}{3}} \sim \frac{mf_\text{eff}}{\sqrt{2}},
\end{equation}
and the TCC constraint (\ref{tcc}) thus requires the parameters of this model to satisfy 
\begin{equation}\label{tccbound_anal}
	\frac{mf_\text{eff}}{\sqrt{2}\omega_+}\ln\left(\frac{f}{A}\right) < \ln\left(\frac{\sqrt{2}}{mf_\text{eff}}\right).
\end{equation}

It is straightforward to check that for a broad range of masses (any $m \lesssim 0.01$) and axion count (any $\mathcal{N} \lesssim 1000$), the constraint (\ref{tccbound_anal}) places an upper bound on $f_\text{eff}$ that is at most $\mathcal{O}(1)$ and asymptotes toward $\sim 0.7$ as the mass decreases. This constraint in the low-mass regime is already somewhat tighter than other Swampland conjectures and string-theoretic arguments, which broadly disfavor scenarios with $f_\text{eff} \gtrsim 1$  \cite{heidenreich_weak_2015, reig_stochastic_2021, agrawal_cosmological_2018, rudelius_possibility_2015}. Importantly, factors of $\mathcal{O}(1)$ in the logarithm on the right-hand side do not meaningfully affect this asymptotic behavior, so our results remain true under slight modifications to the cutoff scales in the TCC.

Due to our choice of initial conditions and analytic approximations, the above constraint on $f_\text{eff}$ is a conservative one. We can produce an even tighter bound on $f_\text{eff}$ by numerically simulating $\mathcal{N}$ axions beginning from rest at precisely $\varphi_n = 0$. To maximize the accuracy of these simulations, we replace the approximations from Eqs. (\ref{approxVp}-\ref{approxH}) with the exact expressions
\begin{equation}
	V'(\varphi_n) = -m^2f\sin(\varphi_n/f)
\end{equation}
and
\begin{equation}
	3H^2 = \sum_{n=1}^N \left[V(\varphi_n) + \frac{1}{2}\dot{\varphi}_n^2\right].
\end{equation}

\begin{figure}
\begin{center}
\includegraphics[width=1.1\columnwidth]{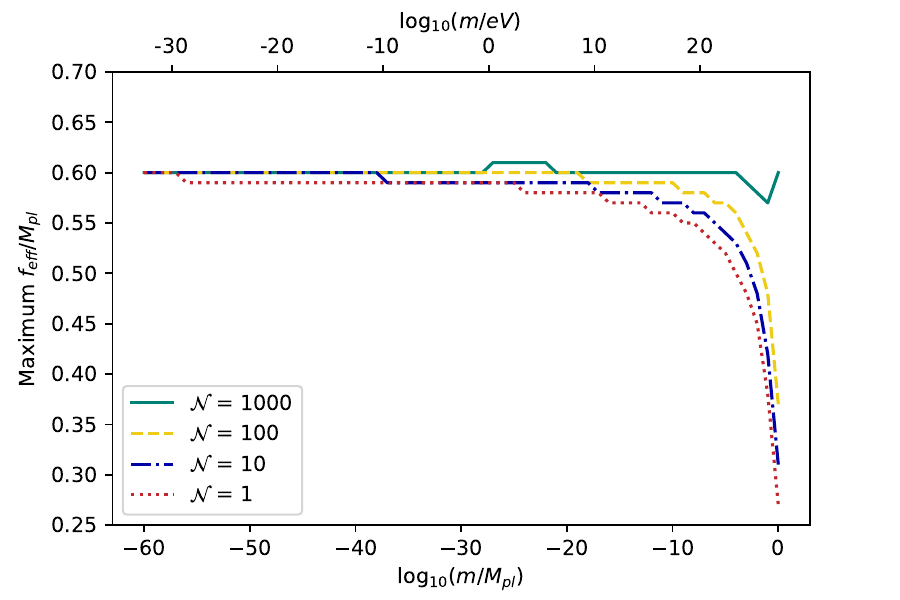}
\caption{\label{fig:fbound}Maximum values of $f_\text{eff} \equiv f\sqrt{\mathcal{N}}$ allowed by the TCC as a function of axion mass $m$ and axion count $\mathcal{N}$, according to numerical simulations with $\pm 0.01M_{pl}$ precision on $f_\text{eff}$. }
\end{center}
\end{figure}

We account for quantum fluctuations in the fields by supplementing their classical evolution (governed by Eq. \ref{eom}) with stochastic jumps in the field values applied independently to each field (see \ref{app} for details). Since the simulations are random in nature, we bound $f_\text{eff}$ by the highest value for which fewer than half of the simulations violate the TCC. This bound is illustrated as a function of $m$ and $\mathcal{N}$ in Fig. (\ref{fig:fbound}), and it closely matches the conclusions from our analytic calculation in the low-mass limit (relevant, e.g., for quintessence models), constraining 
\begin{equation}\label{tccbound_num}
	f_\text{eff} \lesssim 0.6 \quad \text{(for $m \lesssim GeV$)}.
\end{equation}
At higher masses (relevant, e.g., for inflationary models), the constraint on $f_\text{eff}$ is even tighter, reflecting the shrinking hierarchy between the Hubble scale and the Planck scale. The constraints are also generally tighter for systems with lower axion count $\mathcal{N}$, as predicted by the analytic constraint (\ref{tccbound_anal}).

\section{Constraints on Inflation}\label{sec:inflation}
Inflationary models require several features to be successful, including a period of accelerated expansion lasting sufficiently many e-folds to establish a Bunch-Davies vacuum and generate inhomogeneous modes that are just today re-entering the Hubble horizon and are observable in the CMB. These modes must also have an amplitude, tilt, and tensor-to-scalar ratio compatible with observational bounds. In this section, we show that models of axion inflation using the potential (\ref{cosV}) and obeying the TCC constraint (\ref{tccbound_num}) cannot satisfy all of these criteria at once, regardless of initial conditions. 

The modes re-entering the Hubble horizon today were produced $N_e^* \sim 30-60$ e-folds of the scale factor prior to the end of inflation, depending on the reheating temperature \cite{schmitz_trans-planckian_2020}. Note that even protracted reheating scenarios, which have previously been used to adjust $N_e^*$ and improve compatibility between natural inflation and data \cite{stein_natural_2022}, cannot push $N_e^*$ below $\mathcal{O}(30)$ without interfering with Big Bang Nucleosynthesis (BBN). This leads to a constraint on the total number of e-folds of accelerated expansion, $N_e^\text{tot} \equiv \ln(a_\text{end}) \gtrsim 30$. It is straightforward to show that this condition cannot be satisfied in $\mathcal{N}$-axion inflation models with $f_\text{eff} \lesssim 1$ when the fields begin from rest in the lower half of the cosine potential. In this case, the equation of motion (\ref{eom}) is solved by damped oscillations around the minimum scaling roughly as
\begin{equation}
	[\varphi_n(t) - f\pi] \sim e^{-mf_\text{eff}t}\cos(mt) \quad \text{(trough)}.
\end{equation}
It is clear to see that acceleration cannot last for time scales longer than $t \sim m^{-1}$, leading to a severely limited $N_e^\text{tot} = \int Hdt \lesssim f_\text{eff} \lesssim 1$. This rules out inflationary models with initial conditions in the lower half of the potential.

On the other hand, when the axion field values begin near the maxima of their potentials, it is possible to achieve a large number of e-folds while satisfying the TCC bound on $f_\text{eff}$, as long as the axion mass $m$ satisfies 
\begin{equation}\label{tcc_mass}
	mf_\text{eff} \lesssim 10^{-19}.
\end{equation}
This upper limit comes from a model-independent bound on the energy scale of TCC-compliant inflation \cite{bedroya_trans-planckian_2020-1, brandenberger_trans-planckian_2021}. Even if this condition is satisfied, however, one still runs into severe inconsistencies between $\mathcal{N}$-axion inflation models and measurements of the spectral tilt. (For earlier constraints on inflation driven by a single axion, which had already begun to show tensions with observational data, see Refs. \cite{freese_natural_2004, planck_collaboration_planck_2014, pajer_review_2013, akrami_planck_2020}.) 

We know from Eq. (\ref{traj_top}) and the following discussion that each field's trajectory near the hilltop scales roughly as 
\begin{equation}
	\varphi_n(t) \propto e^{mt} \quad \text{(hilltop)}.
\end{equation}
Since the scale factor grows approximately as
\begin{equation}
	a(t) \propto e^{Ht} \approx e^{\sqrt{2/3}f_\text{eff}mt},	
\end{equation}
$N_e^*$ e-folds of the scale factor correspond to about $N_e^*/f_\text{eff}$ e-folds of the field value, implying that $\varphi_n \ll f$ at the time the large-scale CMB modes were being created. Then, since the equation of state (\ref{epsilon}) evaluated at $\varphi_n \ll f$ and $f_\text{eff} \lesssim 1$ is
\begin{equation}
	\epsilon \approx \frac{\varphi_n^2}{f^2},
\end{equation}
the equation of state would have been $\epsilon^* \approx e^{-2N_e^*/f_\text{eff}} \lesssim e^{-60}$ when those CMB modes were created. The tensor-to-scalar ratio $r^* = 16\epsilon^*$ is therefore similarly small and consistent with the observational upper bound \cite{tristram_improved_2022}. On the other hand, the scalar tilt is given by
\begin{equation}
	1 - n_s = \frac{d\ln(\epsilon)}{d\ln(a)} + 2\epsilon \approx \frac{d\ln(\epsilon)}{d\ln(a)},
\end{equation}
and since $\epsilon \propto \varphi_n^2$, we have that
\begin{equation}
	1 - n_s \approx 2\frac{d\ln(\varphi_n)}{d\ln(a)} =  2\frac{d\ln(\varphi_n)}{Hdt} \approx 2f_\text{eff}^{-1},
\end{equation}
up to multiplicative factors of order unity. Clearly, any $f_\text{eff} \lesssim 1$ will be inconsistent with the observed spectral tilt ($1 - n_s \approx 0.03$) by at least two orders of magnitude \cite{akrami_planck_2020}. As a result, even when the number of e-folds is sufficient and consistent with the TCC, models of $\mathcal{N}$-axion inflation obeying Eq. (\ref{tccbound_num}) still fail.

The example of axion inflation considered here illustrates a much more general tension between inflation and the TCC. It has been shown in Refs. \cite{bedroya_trans-planckian_2020-1,brandenberger_trans-planckian_2021} that for TCC-compliant inflationary models to be consistent with the perturbation amplitude observed in the CMB, the equation of state $N_e^*$ e-folds before the end of inflation must satisfy $\ln(\epsilon^*) \lesssim -71$. On the other hand, the observed tilt tells us that $d\ln(\epsilon)/d\ln(a) \sim 0.03$ at that same time. In order to end inflation by achieving $\epsilon = 1$ within $\mathcal{O}(30)$ e-folds of the scale factor, it is necessary for $d\ln\epsilon/d\ln(a)$ to increase \emph{rapidly} to at least $\mathcal{O}(1)$; however, as we have demonstrated, this does not occur naturally on cosine (or any approximately inverse-parabolic) potentials. This issue has been noticed already in previous proposals for TCC-compliant models of inflation \cite{das_swampland_2020, bedroya_trans-planckian_2020-1, schmitz_trans-planckian_2020}. In certain models, it was resolved by a sharp and finely tuned cliff in the inflaton's potential \cite{bedroya_trans-planckian_2020-1} or a finely tuned waterfall phase transition that ends acceleration as soon as the inflaton exits the slow-roll regime \cite{schmitz_trans-planckian_2020}. In any case, it is evident that the TCC requires inflationary models to be supplemented with a ``kill switch'' mechanism that ends acceleration at just the right time to match observational constraints.

\section{Constraints on Quintessence}\label{sec:quintessence}
Unlike inflationary models, a successful model of axion quintessence only needs to achieve $\mathcal{O}(1)$ e-fold of accelerated expansion with a low (quasi-de Sitter) equation of state. Moreover, observational constraints on the time-dependence of the equation of state are significantly weaker than in the case of inflation. As a result, it is possible in principle for axion quintessence models to be simultaneously consistent with the TCC and with all available observational data.

Indeed, it was shown in Refs. \cite{frieman_cosmology_1995, kaloper_pngb_2006, smer-barreto_planck_2017} that models of a single axion with $f \gtrsim 0.5$ can successfully reproduce the behavior of dark energy for a broad range of initial conditions. The primary issue with these models, however, is that they are difficult to construct within string theory. This is because the existence of a supersymmetry breaking sector gives rise to instanton-generated potentials with
\begin{equation}
	V_{SSB}(\varphi) = m_S^2 \cdot e^{-S_{inst}}\cos(\varphi/f),
\end{equation}
where $m_S$ is the scale of supersymmetry breaking and $S_{inst}$ is the instanton action \cite{arvanitaki_string_2010, svrcek_cosmological_2006}. As a result, an axion whose potential is of the same order as the present-day critical energy density must have
\begin{equation}
	m_S^2 \cdot e^{-S_{inst}} \lesssim H_0^2 \implies S_{inst} \gtrsim 2\ln\left(\frac{m_S}{H_0}\right) \gtrsim 200,
\end{equation}
assuming $m_S \gtrsim$ TeV. Because a string axion's decay constant is related to the instanton action via
\begin{equation}\label{lowf}
	f \sim 1/S_{inst} \lesssim 0.005,
\end{equation} 
any string axion sufficiently light to be quintessence will have a decay constant that falls far short of the $f \sim 0.5$ threshold \cite{svrcek_cosmological_2006}.
This line of reasoning suggests that if axions are string-theoretic in origin, one needs at least $N \sim 10^4$ of them to produce a satisfactory effective decay constant, $f_\text{eff} \equiv f\sqrt{N} \gtrsim 0.5$. 

As it happens, depending on the particular Calabi-Yau compactification, models of string theory can contain up to $\mathcal{O}(10^2-10^6)$ light axions in their low-energy limits \cite{mehta_superradiance_2021, svrcek_cosmological_2006}. It is natural to ask whether a collection of this many axions can successfully mimic dark energy if their initial field values are spread in a random uniform distribution across the cosine potential. Unfortunately, numerical simulations show that this scenario would require $f_\text{eff} \gtrsim 1.4$, which violates the TCC by at least $\mathcal{O}(\ln(m^{-1}))$ e-folds of accelerated expansion.

\begin{figure}
	\begin{center}
		\includegraphics[width=\columnwidth]{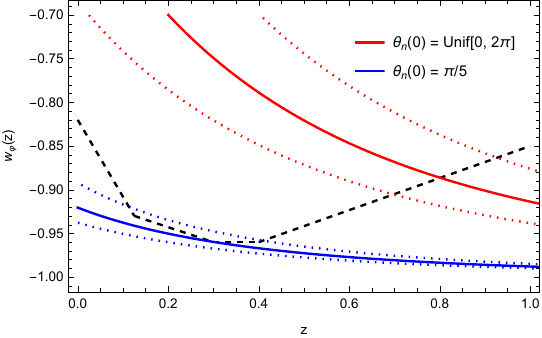}
		\caption{\label{fig:quintessence}
		Comparison of theoretical predictions (solid/dotted curves) to the observational upper bound (dashed black curve, adapted from Ref. \cite{montefalcone_dark_2020}) on the dark energy equation of state $w_\varphi$ as a function of redshift $z$. The solid-curve theoretical predictions are generated under the assumption of $N \gtrsim 10^4$ identical axions with $f_\text{eff} = 0.6$, while the dotted curves illustrate sensitivity to $f_\text{eff} \in \{0.55, 0.65\}$. The blue curve, which is consistent with the observational constraints, assumes all of these axions have initial field values satisfying $\theta_n(0) \equiv \varphi_n(0)/f = \pi/5$, while the red curve, which violates observational constraints, assumes the initial positions are randomly, uniformly distributed across their domain. Note that these theoretical predictions are generated by numerically simulating the axions' classical evolution, starting from zero initial velocity at early times ($z \gg 1$), in the presence of ordinary (dust-like) matter. Present-day ($z=0$) is defined by reaching the fractional energy densities $\Omega_m = 0.3$ and $\Omega_\varphi = 0.7$.}
	\end{center}
\end{figure}

In Fig. (\ref{fig:quintessence}), we compare the dark energy equation of state $w_\varphi \equiv P_\varphi/\rho_\varphi$ as a function of redshift $z$ for a scenario with a random uniform distribution of initial field values (red curve) versus one with identical field values near the hilltop (blue curve), each with $f_\text{eff} = 0.6$. Only the model with initial field values near the hilltop is compatible with the observational upper limit on $w_\varphi$.
While requiring that each initial field value satisfies $|\theta_n(0)| \equiv |\varphi_n(0)|/f \lesssim \pi/5$ would be a modest constraint for single-axion models, it is a much more demanding one for models with $N \gtrsim 10^4$ axions. This result suggests that mechanisms for dynamically positioning axions near the hilltop, such as the maximal-misalignment mechanism used in \emph{N-essence} \cite{reig_stochastic_2021}, may be essential for models of string-axion quintessence.

If the axions do not saturate the TCC bound and instead have a lower $f_\text{eff}$, it is still possible to comply with observational constraints, but only for more finely tuned initial values of $\varphi_n$. We can estimate the necessary tuning by adapting the result from Eq. (\ref{nend}) to variable initial conditions $\{\varphi_n(0),  \; \dot{\varphi}_n(0)\}$ and taking the limit $f_\text{eff} \ll 1$, finding that the number of e-folds of acceleration accrued during classical evolution is given by
\begin{equation}\label{nend2}
	N_e^\text{tot} \approx f_\text{eff} \ln\left(\frac{2f}{\varphi_n(0) + \dot{\varphi}_n(0)/m}\right).
\end{equation}
The argument of the logarithm depends on whether the axion begins in slow-roll\footnote{We caution that even if the axion is slowly rolling at the onset of acceleration, much of its later trajectory will occur \emph{outside} of the slow-roll regime. Indeed, one can check using Eqs. (\ref{traj_top}) and (\ref{omega}) that a necessary condition for slow-roll, $\ddot{\varphi} \ll H\dot{\varphi}$, breaks down when the growing mode is dominant in models with  $f_\text{eff} \lesssim 1$.} (with $\dot{\varphi}_n(0)/m \sim \varphi_n(0)/f_\text{eff}$) or with negligible field velocity ($\dot{\varphi}_n(0)/m \ll \varphi_n(0)$), though this difference will not be too important.
In order to achieve at least $\mathcal{O}(1)$ e-fold of accelerated expansion, $\varphi_n(0)$ must satisfy 
\begin{equation}\label{tuning}
	\varphi_n(0)/f \lesssim e^{-f_\text{eff}^{-1}},
\end{equation}
where we have neglected the sub-exponential scaling with factors dependent on the initial field velocity. Note that in the limit of a single axion, this upper bound reproduces the result from Ref. \cite{kaloper_pngb_2006} and roughly represents the probability that an axion with a random initial field value can drive quintessence. This bound differs from the probability found in Ref. \cite{kamionkowski_dark_2014} for two important reasons. First, we used the full definition of the equation of state (Eq. \ref{epsilon}) to determine when acceleration ends, rather than the approximate slow-roll expression $\epsilon \approx \frac{1}{2} |\nabla V/V|^2$.  Second, we required acceleration to continue for $\mathcal{O}(1)$ e-fold to satisfy observational constraints for quintessence, whereas the limits in Ref. \cite{kamionkowski_dark_2014} included cases where the duration of accelerated expansion is arbitrarily shorter.

From Eq. (\ref{tuning}), we see that extreme fine-tuning can classically compensate for arbitrarily low values of $f_\text{eff}$ in axion quintessence models. In practice, however, quantum fluctuations can destabilize extremely fine-tuned configurations. Whether fluctuations in the field are on the order of some TCC-compliant inflationary energy scale, the present Hubble scale, or even several orders of magnitude lower, they impose a lower bound on $\varphi_n(0)/f$ and in turn constrain
\begin{equation}
	f_\text{eff} \gtrsim 0.01.
\end{equation} 
This bound is slightly looser than the analogous calculation for a single axion in Ref. \cite{kaloper_pngb_2006} (which assumed a higher, TCC-violating inflationary energy scale), but it still disfavors models of quintessence driven by a single string axion with $f \lesssim 0.005$.

Finally, we comment on the possibility of a nonzero cosmological constant and its effect on axion quintessence models. Obviously, if the vacuum energy density is small and positive, there is no need for quintessence in the first place. However, vacua in string theory have a notorious preference for negative energy densities \cite{juanmaldacena_supergravity_2012}, and explicit models have been constructed where the negative vacuum energy density $\rho_\text{vac}$ is smaller in magnitude than the present dark energy density $\rho_{DE} \sim H_0^2$ \cite{demirtas_exponentially_2022}. In the limit $|\rho_\text{vac}| \ll \rho_{DE}$, the phenomenology of axion quintessence models would be indistinguishable from that arising in the present work under the assumption $\rho_\text{vac} = 0$. Moreover, numerical calculations show that assuming a more comparable vacuum energy density, such that $V(\varphi_n) = m^2f^2\cos(\varphi_n/f)$ with no constant offset, would only change the constraints on $f_\text{eff}$ (calculated in Section \ref{sec:tcc}) and $|\theta_n(0)|$ (calculated in this section) by $\mathcal{O}(10\%)$. These conclusions are consistent with Ref. \cite{ruchika_observational_2023}, which found axion quintessence models to be compatible with the presence of a small cosmological constant of either sign.

\section{Conclusions \& Discussion}\label{sec:conclusions}
The central finding in this work is that the TCC constrains any system of $\mathcal{N}$ identical axions with simple cosine potentials to have decay constants satisfying $f_\text{eff} \equiv f\sqrt{N} \lesssim 0.6$ in reduced Planck units. This bound is even tighter for axions with masses near the Planck scale (see Fig. \ref{fig:fbound}). Because the TCC must hold for any physically allowed initial conditions (and not just the initial conditions in our own observable universe), these constraints apply to all such systems of axions, regardless of whether or not they are responsible for driving accelerated expansion.

We have shown that this constraint rules out models of axion-driven inflation, as larger values of $f_\text{eff}$ are required to achieve sufficiently many e-folds of inflation and produce the correct spectral tilt. 
We have also argued that reconciling \emph{any} inflationary model with the TCC---axionic or otherwise---requires a mechanism for ending inflation via a sharp and sudden increase of the equation of state $\epsilon$, rather than the traditional graceful exit. Some examples in the literature show that this can be accomplished in principle, but they rely on an extraordinary amount of fine-tuning \cite{bedroya_trans-planckian_2020-1, schmitz_trans-planckian_2020}.

In contrast to axion-driven inflation, models of axion quintessence \emph{can} be simultaneously compatible with the TCC and observational data, as long as $0.01 \lesssim f_\text{eff} \lesssim 0.6$ and the axions' initial field values are near the top of the potential at the level of tuning specified by Eq. (\ref{nend2}). In the case of a single axion with $f \approx 0.6$, the necessary tuning is relatively modest, requiring the initial dimensionless field value to satisfy $|\theta| \equiv |\varphi|/f \lesssim \pi/5$. Ultralight string axions, however, typically have much lower decay constants $f \lesssim 0.005$, and one would therefore need $\mathcal{N} \gtrsim 10^4$ of them, \emph{each} aligned to within $\pm \pi/5$ radians of the maximum, in order to achieve the same effect. Alternatively, one could have fewer string axions (resulting in a lower $f_\text{eff}$) with more finely tuned initial conditions, up to the limit set by quantum fluctuations. In either case, axions that are specifically string-theoretic in origin appear to require a mechanism to perch them near the hilltop.

We emphasize that the bounds and constraints derived in this work apply specifically to models of $\mathcal{N}$ identical axions with cosine potentials. Models of inflation making use of multiple non-identical axions \cite{czerny_multi-natural_2014, bachlechner_chaotic_2015} or axions coupled to a bath of radiation \cite{visinelli_natural_2011, mishra_warm_2012, montefalcone_observational_2023} are not directly constrained by the present work, but they are unlikely to overcome the general obstacles for TCC-compliant inflation outlined above. Models of quintessence using axions with a small range of masses and decay constants would likely still be feasible with some additional tuning, but more elaborate models of axion quintessence, which may incorporate nontrivial interactions between axions, monodromies, contributions from higher-order instanton corrections, or interactions with dynamical moduli, require their own independent analysis.

\section{Acknowledgements}
I wish to thank Paul Steinhardt for providing guidance during this study and suggestions for the manuscript. I am also grateful to Alek Bedroya, Anna Ijjas, and Anirudh Prabhu for their helpful feedback on the manuscript. This work was supported in part by the DOE grant number DEFG02-91ER40671 and by the Simons Foundation grant number 654561.

\appendix
\section{Quantum Fluctuations}\label{app}
While the axion fields are located at the hilltops of their potentials with sufficiently low classical velocities, their evolution may be dominated by quantum fluctuations. The initial conditions used for the analytic calculation in Section \ref{sec:tcc}, namely $\varphi_n(0) = \frac{mf_\text{eff}^2}{\pi\sqrt{6}}$ and $\dot{\varphi}_n(0) = \frac{m^2f_\text{eff}}{6\pi}$, are specifically chosen to avoid this regime. To see this, we can consider the RMS fluctuation \cite{bardeen_quantum_1987} of a massive field in a de Sitter background,
\begin{equation}\label{rms}
	(\Delta \varphi_n)_\text{rms} = \sqrt{\frac{H^2}{8\pi^2\eta}(e^{2\eta\ln(a)} - 1)},
\end{equation}
and compare its time-derivative
\begin{equation}
	\dot{(\Delta \varphi_n)}_\text{rms} = \frac{H^3e^{2\eta\ln(a)}}{8\pi^2(\Delta \varphi_n)_\text{rms}}
\end{equation}
to the field's classical velocity. Here, $a$ is the scale factor, $\eta = -m^2/(3H^2)$, and we have taken $H$ to be approximately constant. Since $\eta < 0$, we have that
\begin{equation}
	\dot{(\Delta \varphi_n)}_\text{rms} < \frac{H^3}{8\pi^2(\Delta \varphi_n)_\text{rms}},
\end{equation}
so we can conservatively estimate that $\dot{(\Delta \varphi_n)}_\text{rms} \lesssim \dot{\varphi}_n(0)$ when 
\begin{equation}
	\frac{H^3}{8\pi^2(\Delta \varphi_n)_\text{rms}} \lesssim \frac{m^2f_\text{eff}}{6\pi} \iff (\Delta \varphi_n)_\text{rms} \gtrsim \frac{mf_\text{eff}^2}{\pi\sqrt{6}}.
\end{equation}
In other words, the initial conditions we chose ensure that quantum effects offset the initial field values by a factor less than $\mathcal{O}(1)$. Additionally, the classical equations of motion (\ref{eom}) ensure that $\ddot{\varphi}_n \approx m^2\varphi_n - 3H\dot{\varphi}_n$ remains positive when starting from these initial conditions, amplifying the fields' classical velocities while $\dot{(\Delta \varphi_n)}_\text{rms}$ falls off.

To simulate the effects of quantum fluctuations numerically, we employ a random walk with time step $dt$, where at each time step, the field value of each axion changes by
\begin{equation}
	\delta\varphi_n = \pm \frac{H}{2\pi}\sqrt{Hdt} \cdot e^{\eta\ln(a)}.
\end{equation}
At each time step, the scale factor changes according to $d\ln(a) = Hdt$, and so the variance of this random walk at some future time with scale factor $a$ will be, indeed,
\begin{align}
	\langle\varphi_n^2\rangle &= \sum_{j=0}^{\frac{\ln(a)}{Hdt}} \frac{H^3}{4\pi^2}dt \cdot e^{2\eta (jHdt)} \\
	&\approx \int_{j=0}^{\frac{\ln(a)}{Hdt}} \frac{H^3}{4\pi^2}dt \cdot e^{2\eta (jHdt)}dj \\
	&= \frac{H^2}{8\pi^2\eta}(e^{2\eta\ln(a)} - 1).
\end{align}

\bibliographystyle{elsarticle-num.bst}
\bibliography{axions.bib}

\end{document}